# A Survey on Human and Personality Vulnerability Assessment in Cyber-security: Challenges, Approaches, and Open Issues

DIMITRA PAPATSAROUCHA*, YANNIS NIKOLOUDAKIS, IOANNIS KEFALOUKOS, EVANGELOS PALLIS, and EVANGELOS K. MARKAKIS,

Department of Electrical and Computer Engineering, Hellenic Mediterranean University

These days, cyber-criminals target humans rather than machines since they try to accomplish their malicious intentions by exploiting the weaknesses of end users. Thus, human vulnerabilities pose a serious threat to the security and integrity of computer systems and data. The human tendency to trust and help others, as well as personal, social, and cultural characteristics, are indicative of the level of susceptibility that one may exhibit towards certain attack types and deception strategies. This work aims to investigate the factors that affect human susceptibility by studying the existing literature related to this subject. The objective is also to explore and describe state-of-the-art human vulnerability assessment models, current prevention, and mitigation approaches regarding user susceptibility, as well as educational and awareness-raising training strategies. Following the review of the literature, several conclusions are reached. Among them, Human Vulnerability Assessment has been included in various frameworks aiming to assess the cyber security capacity of organizations, but it concerns a one-time-assessment rather than a continuous practice. Moreover, human maliciousness is still neglected from current Human Vulnerability Assessment frameworks; thus, insider threat-actors evade identification, which may lead to an increased cyber security risk. Finally, this work proposes a user susceptibility profile according to the factors stemming from our research.

**Keywords**: human vulnerability assessment; personality assessment; personality measures; cyber-security; cyber-security training; social engineering.

## 1 INTRODUCTION

Continuous technological advancements demand the establishment of secure systems and the development of defense mechanisms against malicious actors. These tasks prove to be very complex due to various factors that need to be taken into consideration (e.g., technological aspects, attack patterns, attacker and defender behavior, etc.). Cyber-security is the broad field that deals with these challenges. Attacks in the cyber-space may threaten integrity, confidentiality, and availability of computer systems, networks, software programs, and data. As described in [8,37,90] cyber-security includes technologies, security policies, defense mechanisms, and mitigation strategies for:

- identifying and managing cyber-security risk of data, systems, assets, and capabilities
- preventing cyber-attacks and protecting systems, networks, software, data, and eventually individuals
- detecting cyber-attacks
- responding to cyber-security incidents and counteracting against cyber-attacks
- recovering from cyber-attacks

*Authors' addresses: D. Papatsaroucha (corresponding author), Y. Nikoloudakis, I. Kefaloukos, E. Pallis, and E. Markakis, Department of Electrical and Computer Engineering, Hellenic Mediterranean University, 71410 Heraklion, Greece; emails: d.papatsaroucha@pasiphae.eu, nikoloudakis@pasiphae.eu, g.kefaloukos@pasiphae.eu, pallis@pasiphae.eu, markakis@pasiphae.eu.

Lately, there has been an ongoing investigation concerning the factors that affect crime rate in general, in an endeavor to gain insight about what shapes illegal activity. Bothos and Thomopoulos in [22] utilized several econometric approaches to address this matter regarding a large urban environment. Results of their study showcased that school dropout and poverty are factors that play an important role in crime rate increase, while effective law enforcement may result in a decrease.

Malicious activity in the cyber-world is a type of crime that is usually referred to as cyber-crime. Alongside the general rise in crime rates, a rise in cyber-crime has been also noticed, with the following factors contributing to this increase. Firstly, most cyber-security approaches target technological vulnerabilities [79,91], resulting in the development of solutions for patching them; however, many researchers of the cyber-security field agree that the human factor appears to be the weakest link in cyber-security, while highly vulnerable users are often referred to as "weak-link users" [3,10,11,48,77,96,112,114,127]. Secondly, according to Conteh and Roger in [34], the attackers' ability to retain their anonymity and the defenders' lack of digital literacy are only some of the reasons that intrigue cyber-attackers to engage in malicious activities. Additionally, the number of potential intrusion points, which can be exploited by cyber-criminals, has expanded. That is due to various reasons such as the growth in internet connectivity and internet connected devices and services (Internet of Things – IoT / Internet of Everything - IoE), and the increase in the number of social media users the past few years. Statistical analysis has revealed that the latter has increased to more than half of the global population, by the start of 2021[*].

## 1.1 Social Engineering (SE)

An increasing number of cyber-attacks that aim to exploit the vulnerabilities of the human factor has been observed recently. These attacks constitute a broad category called social engineering (SE). Accordingly, in this paper, the cyber-attackers that perform SE attacks will be referred to as "SE attackers". SE is not at all a new concept, as it has its roots in the ageless art of fraud and deception. However, it is the "naive" nature of humans and the extensive use of social media that has elevated this technique to a current trend [12]. Through SE techniques, a malicious actor (human or program) aims to convince potential victims to proceed with certain actions (e.g., disclosing confidential information, providing access to restricted locations – physical or digital, etc.) by exploiting their vulnerabilities [86].

Research has shown that in 2020, 70% of data breaches were due to credentials' theft and 22% of them were caused by SE attacks [89]. Moreover, 96% of SE attacks were launched via email (phishing). Phishing, which is one of the most popular SE attacks, ranked first in the list with the 16 more common attacks related to data breaches, and second in the list with the 16 more common attacks [89].

---

[*] https://datareportal.com/reports/digital-2021-global-overview-report

According to Conteh and Royer in [34], false memories (distorted recollection of events or completely fabricated memories [23]), difficulty to stay concentrated, and psychological weaknesses, like the tendency to trust and help others, are only some of the reasons that humans tend to be more susceptible to SE attacks than to any other type of cyber-attacks. What is more, SE attacks may cause several implications to organizations, such as financial and reputational damage, time investment for loss recovery, etc. Hence, the authors of [34] underlined that it is mandatory to address these kinds of attacks appropriately with regular, thorough, and relevant awareness-raising training of employees. Additionally, the authors proposed the performance of frequent penetration tests and SE drills, and the establishment of organizational disciplines regarding security policies.

### 1.2 Human and Personality Vulnerability Assessment (HVA, PVA)

The aforementioned observations have created a relatively new focal point in the research on cyber-security, the assessment of human vulnerabilities. Several personality traits such as agreeableness and neuroticism are considered to be indicative of susceptibility to cyber-threats [65]. Other personal characteristics, such as one's cognitive processes (e.g., information-processing, decision-making, etc.) have concerned researchers of the field, regarding their influence on cyber-security intentions and behavior, as well as in the ability to detect frauds in general. Furthermore, when trying to identify potential vulnerabilities, one should also take into consideration several other factors that may influence security-related behavior, such as users' demographics, users' habits, etc.

### 1.3 Aim of Research and Paper Structure

The assessment of human vulnerabilities is an essential aspect of cyber-security. Hence, research needs to be stirred towards the human factor for delivering complete security solutions. Researchers and professionals of the field investigating human vulnerabilities need to take into account many variables (e.g., psychological), hence the assistance of social scientists, psychologists, and human-factor experts[*] is critical. Moreover, although preserving cyber-security is important for systems and data of organizations and industries, it is also important for human welfare, as personal privacy is at stake. It is pivotal for people to be properly educated about how to protect themselves in the cyber-space, recognize potential threats, and confront them, whether it affects their professional or personal life. Therefore, in an endeavor to gain insight into these issues, the goal of this review is to investigate the existing literature and gather information about the following:

- factors influencing individual susceptibility to cyber-attacks, such as SE
- human characteristics (personality related or not) that are indicative of susceptibility and that will be used for the development of the proposed user susceptibility profile
- existing frameworks and methodologies that aim to assess human vulnerabilities and susceptibility to cyber-threats
- existing tactics for awareness-raising training

This paper also attempts to reach some conclusions based on the literature research, by identifying any limitations of the state-of-the-art and propose possible future steps and directions regarding research and assessment or mitigation approaches related to human vulnerabilities in the context of cyber-security.

---

[*] Human-factor experts or researchers are those who have studied the ways that humans relate to technology and aim to improve the experience of end-users in order to achieve higher operational performance and enhance safety.

The rest of the paper is structured as follows: Section 2 elaborates on our motivation to conduct this research; Section 3 provides a brief presentation of the basic concepts described in this paper (personality traits, cognitive processes, inventories, etc.); Section 4 refers to the effect of the described concepts – of Section 3– in shaping human vulnerabilities, while it also refers to current solutions in HVA; Section 5 explores educational and awareness-raising tactics, as well as their results. Finally, Section 6 presents the Conclusions stemming from this literature review.

## 2 MOTIVATION

SE is a technique that in the past was performed only by extremely skilled and talented individuals [12]. However, today it seems to be a common and preferred approach by many cyber-offenders [12]. Technology is developing rapidly, and smart devices are becoming a part of our everyday life. Smart cars and smart houses are no longer a science fiction scenario, and remote working has increased more than ever before, due to the Covid-19 pandemic[*]. This kind of evolution generates more potential intrusion points daily and turns the cyber-world into a playground for malicious actors and cyber-criminals [34]. The following subsections aim to describe the leading role of human vulnerability in everyday cyber-security incidents and the major implications that may be caused due to susceptibility to cyber-threats; the motivating factors that led to the conduction of this survey.

### 2.1 Phishing

Phishing is one of the major SE tactics, through which malicious actors seek to convince the victim to proceed to risky actions (e.g., disclose sensitive information like user credentials or bank account information; click on malicious URLs, etc.). Legitimate but vulnerable websites can be compromised by cyber-criminals and used as phishing websites, as they offer attackers the advantage of not having to deal with domain registration and hosting services for achieving their goals (i.e., launching other malicious actions). This issue concerned Corona *et al.* in [36], who developed a system able to distinguish compromised websites from legitimate ones with an accuracy rate of 99%. DeltaPhish, which is the name of the system, can be installed alongside the firewall of a website and is also able to detect sophisticated attacks that protract traceability, with 70% accuracy. On the other hand, Parsons *et al.* in [98] researched human susceptibility to phishing emails. The authors found out that most of the subjects that took part in their research were unable to identify a genuine from a phishing URL.

Studies like the one mentioned above are called phishing studies and they investigate and elaborate victimization in phishing attacks. Busch *et al.* in [25] introduced a long-term study about the ethical implications and consequences that phishing studies may have. Their study focused on recruiting a large number of participants from several organizations for observing possible short- and long-term, negative, emotional, and social effects of deception. However, this is an ongoing study, and its results will be announced in the future. According to the authors, this study will shed light upon ethical guidelines for researchers, as well as countermeasures and interventions that can be adopted after the conduction of phishing studies, for addressing such effects. Moreover, a study like this will also point out implications caused by deception on a personal level and the effects of exploitation in personal and professional life of victims, which has not been explored adequately yet.

---

[*] https://remoters.net/remote-work-trends-future-insights/

## 2.2 Open Data

The term open data refers to publicly, online available information that can be distributed, shared, and used without restrictions [14]. On that basis and considering the increased popularity of data science techniques in SE attacks, Mauri *et al.* in [81] explored the possibility of open data being involved in an SE attack scenario. The authors also investigated to what extent end-users are adequately informed about security and privacy issues concerning open data. Their hypothetical scenario involved the open data of the University of Cagliari and was based on Mitnick's four-phase attack model [84]:

- research (collecting information)
- development of a good relationship with the victim based on trust (best combination of SE skills)
- exploitation of trust (interaction with the victim)
- utilization of information collected by the attack (i.e., use them to perform another attack).

According to their conclusions, the presented attack scenario, which is very likely to happen in many universities, is very close to reality, easy to automate, and thus a more secure way for data sharing needs to be established. Moreover, the authors of [81] proposed that users need to get informed about cyber-security threats and trained on how to retain their safety in the cyber-space.

## 2.3   Internet of Everything (IoE)

Previous research has revealed that people mostly interact with social media and networks through their mobile devices and therefore they are very likely to experience an SE attack through them [38]. People living in smart IoT-based homes often use mobile devices for handling all the different applications and sensors. Ali and Awad in [6] utilized the OCTAVE (Operationally Critical Threat Asset and Vulnerability Evaluation) Allegro Methodology to conduct a risk assessment in IoT-based homes, which revealed both technical and human vulnerabilities. In case of identity and credential theft, an attacker can impersonate the legitimate user and gain access to the main system of the smart home. This may result in several other attacks, such as sensor undermining, which may further lead to information loss, financial loss, etc. The authors also proposed that constant training and education of users, to gain awareness concerning SE, can assist in preserving privacy and security of IoT-based smart homes.

Moreover, human vulnerabilities also seem to be at the center of attention regarding security of autonomous vehicles [60]. Users often use their smartphones to access their autonomous vehicles, thus facing similar SE threats as people living in IoT-based homes. What is more, developing secure user authentication solutions for autonomous vehicles is a task with high complexity. That is because there is a need for safe, but quick and easy access that will also consider individuals with disabilities (e.g., visually impaired people). According to Linkov *et al.* in [77], human-factor researchers [131] can assist in achieving this manifold goal. Moreover, according to the authors, the faulty or uneducated way a user may respond to an ongoing cyber-attack is also considered a human vulnerability, which may lead to further repercussions. This is because people tend to act impulsively and randomly under stress and time pressure.

## 2.4 User Maliciousness

Another important aspect of human-factor vulnerabilities is maliciousness. Linkov *et al.* in [77] highlighted that the employees of companies that handle autonomous vehicles need to be carefully selected and trusted. That is because of the likelihood that an employee would act as an insider threat and jeopardize the safety of the autonomous vehicle. While human vulnerabilities and susceptibility to SE have attracted the interest of researchers, little research has been conducted about human maliciousness in the field of information security. To this day, many cyber-offenders remain anonymous and evade arrest, thus the study of maliciousness and cyber-attackers' behavior retain a high complexity [34]. Human maliciousness can be a serious human vulnerability for organizations and industries. Malicious or prone to maliciousness employees may act as inner intruders by intention or under certain triggering circumstances that offer them the opportunity to act maliciously [73]. Research has shown that in 2020 internal actors were responsible for 30% of the recorded data breaches [89].

## 3 THEORETICAL BACKGROUND

This section aims to describe briefly, but concisely, the following concepts that influence human cyber-security behavior, which require a deeper knowledge of some basic concepts of psychology:

- personality traits
- cognitive processes (i.e., information-processing, decision-making, risk-taking, and protection motivation)
- security-related behavior intentions
- persuasion techniques (commonly utilized by SE attackers to convince their victims to proceed with certain actions [53])

### 3.1 Personality Traits

People differ from one another in many ways. Personality traits are patterns that reflect the way one may think, feel, and behave [50]. These patterns describe one's personality and each person may present a low, medium, or high level of each one of these traits [129]. Personality traits are shaped by several different factors, such as genetics and experiences, as well as influences, both social and environmental [129]. They can also be strong indicators of the possibility a person may get involved in malicious actions or engage in risky activities [93], as well as of one's security-related behavior intentions [56].

#### 3.1.1 Five-Factor Model (FFM)

One of the most widely accepted and used personality models is the Five Factor Model (FFM), which includes five distinct personality traits that have been found to be unalterable across age and culture groups, and consistent through time [46]. The FFM comprises the following personality traits [66,105,130]:

- *Openness*, which refers to an individuals' tendency to be open-minded towards new ideas and experiences, and to accept different beliefs. People who score high at the *openness* trait exhibit a high appreciation of art, increased imagination, and eagerness for adventure. On the other hand, people who score low feel more comfortable in their routine and do not seek new experiences.
- *Conscientiousness*, which includes characteristics like honesty, trust, strong self-orientation, and self-responsibility. People who score high on this trait are more likely to stick to plans and follow rules.
- *Extraversion*, which relates to social skills. People scoring high on this trait feel comfortable in large groups of people and tend to be enthusiastic, energetic, and talkative. On the contrary, people scoring low on this trait can be described as *introverts*, and therefore, they may feel more comfortable in smaller groups of people.
- *Agreeableness*, which describes people who are more likely to help and trust others because they always assume the best in other people. Depending on how high or low an individual scores on this trait, agreeableness can be a measure of kindness and compassion.

- *Neuroticism*, which refers to increased levels of anxiety. The higher someone scores on this trait, the more one tends to worry, while lower scores indicate emotional stability.

In Table I the personality traits of the Five Factor Model (FFM) are briefly depicted, as illustrated in [54,58].

Table I Personality Traits of the FFM

| Personality Trait | High | Low |
|---|---|---|
| Conscientiousness | Organized, reliable, thorough | Unreliable, negligent, careless |
| Agreeableness | Trusting, kind warm | Selfish, hostile, distrusting |
| Neuroticism | Moody, nervous, tamperamental | Steady, calm |
| Openness | Curious, imaginative | Shallow, conservative |
| Extraversion | Active, assertive talkative | Passive, reserved, quiet |

Using the Big Five Inventory (BFI)[*], the personality traits of an individual can be measured through a questionnaire of 50 questions [66]. Personality traits can serve as indicators of the cyber-security-related behavior intentions of users regarding their computer devices. In [48] it was revealed that certain personality traits, like agreeableness, showcase statistically significant relationships with victimization. On the other hand, in the same study, neuroticism was found to be associated with computer anxiety (i.e., anxiety generated when using a computer device) thus people with this personality trait have been recognized as greatly concerned about their security and privacy (security worries), which may decrease susceptibility to SE attacks, like phishing.

### 3.1.2 Myers-Briggs Type Indicator (MBTI)

The personality traits defined in the FFM are not the only characteristics that may describe feelings, thoughts, and behaviors of humans. As mentioned before, several types of personality scales exist, with Myers-Briggs Type Indicator[*] (MBTI) being among them. As stated in the MBTI [39], people differ in the way they prefer to focus their attention (*extroverts / introverts*), the way they gather information (*sensing / intuition*), the way they come to decisions or make judgments (*thinking / feeling*), and the way they choose to live their lives (*judging / perceiving*).

---

[*] Big Five Inventory – Questionnaire
https://www.ocf.berkeley.edu/~johnlab/bfi.htm
[*] MBTI - https://www.mbtionline.com/en-US

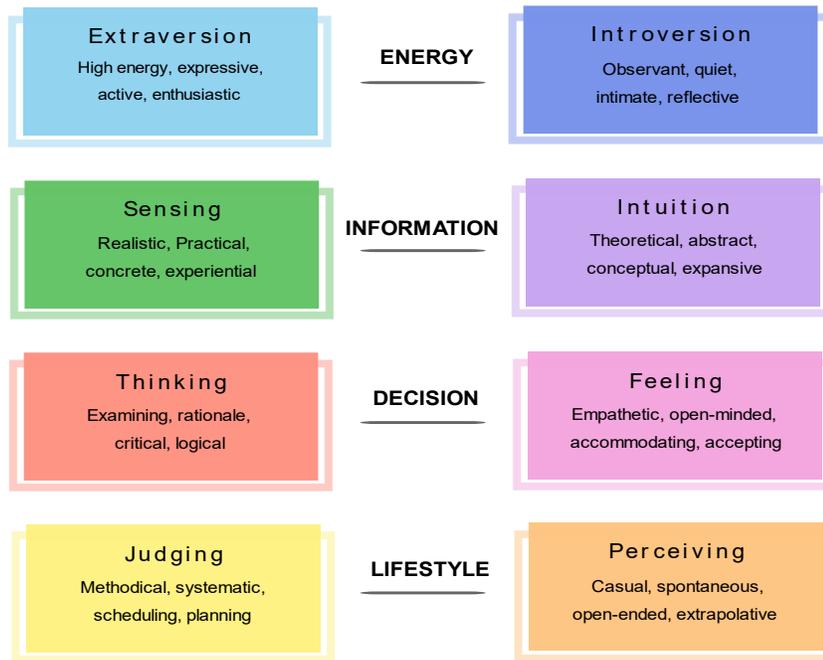

Figure 1. Personality Traits of the Myers-Briggs Type Indicator (MBTI)

According to Isabel Briggs Myers in [39], *extroverts* focus on the external environment, while *introvert*s are more concerned about the inner world. *Sensing* people tend to gather information through their senses and focus on what is real, whereas people driven by *intuition* in information gathering prefer to trust their instincts. Regarding decision-making and judgment, *thinking* people are more objective and rely on logic for drawing conclusions. On the contrary, *feeling* people judge under the guidance of their emotions and exhibit an increased level of empathy. Lastly, one's lifestyle can either be well planned and organized, and this is attributed by the *judging* trait, or it can be more open to possibilities like the *perceiving* trait dictates [88]. Figure 1 shows all traits of the Myers-Briggs Type Indicator (MBTI) alongside their counterparts (as illustrated by the Myers & Briggs Foundation[*]), while in Figure 2 a mapping between the personality traits of the FFM and the MBTI is presented, according to [51]. As one may observe in Figure 2, there is no clear mapping between the neuroticism personality trait of the FFM and any of the personality traits of the MBTI.

### 3.1.3 Dark Triad

Another personality model is the Dark Triad, which aims to explore the darker side of human personality and focuses on characteristics that indicate a tendency to manipulate and deceive [68]. The three traits described in this model are *Psychopathy*, *Narcissism*, and *Machiavellianism*.

---

[*] MBTI - https://www.myersbriggs.org/

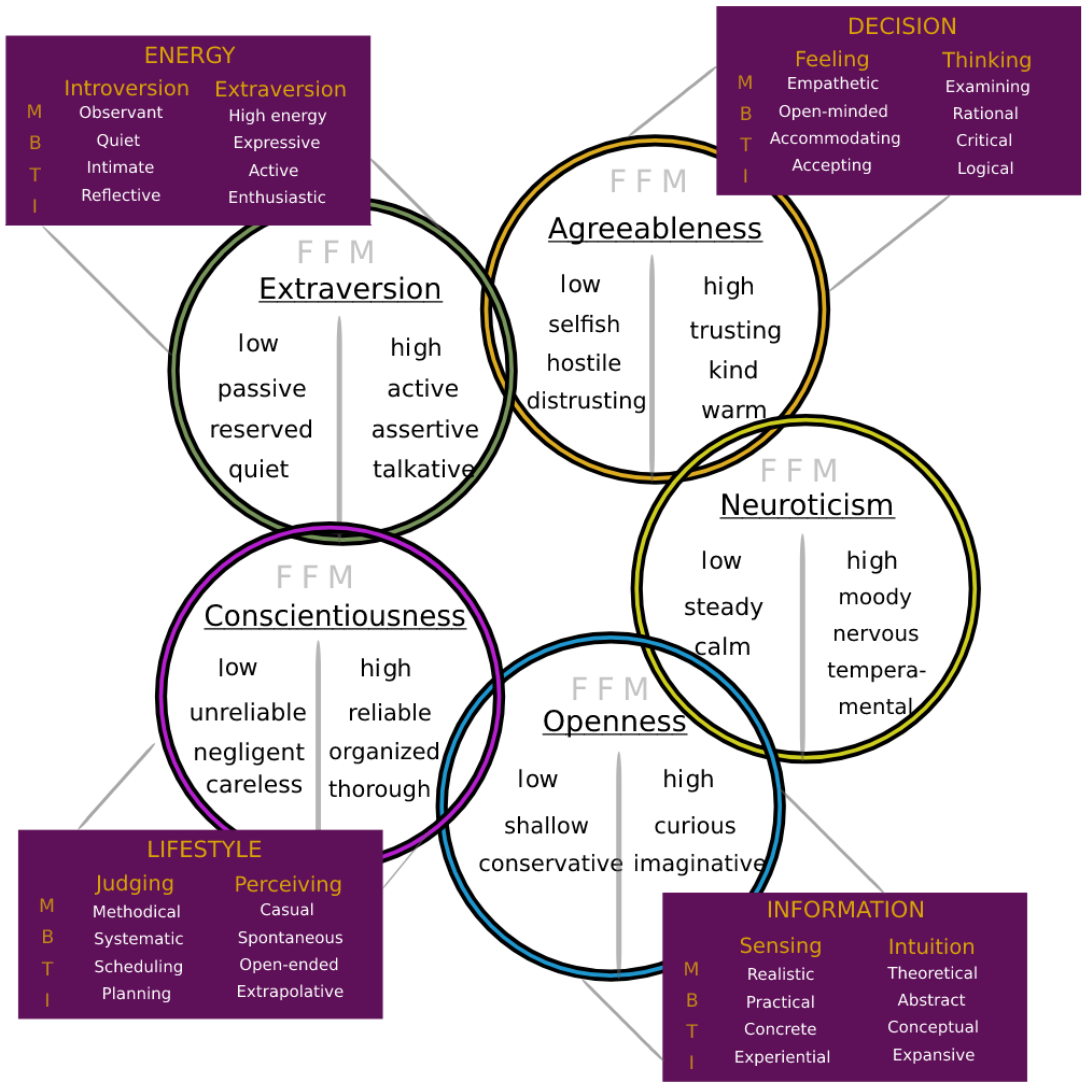

Figure 2. FFM and MBTI Personality Traits Mapping

*Psychopathy* refers to the lack of empathy, while people scoring high on this trait appear to be excellent manipulators, as they acquire the ability to imitate socially accepted behaviors [15,20,125]. People scoring high on *machiavellianism* exhibit lower levels of morality when coming down to increasing personal gain, which may also lead to a tendency for manipulation [18], [30]. An increased level of *narcissism* indicates a person with an idealized image of self, an egoistic behavior, and an excessive need for attention, which as a result makes these individuals less empathetic towards others [94], [45]. Several instruments can be utilized for measuring the Dark Triad personality traits. Some of them measure one trait at a time [83], while there are instruments such as the Dirty Dozen [67] and the Short Dark Triad (SD3)* [101], which measure all three traits at once; hence, they appear to be the most commonly used lately. Figure 3 illustrates the personality traits that constitute the Dark Triad.

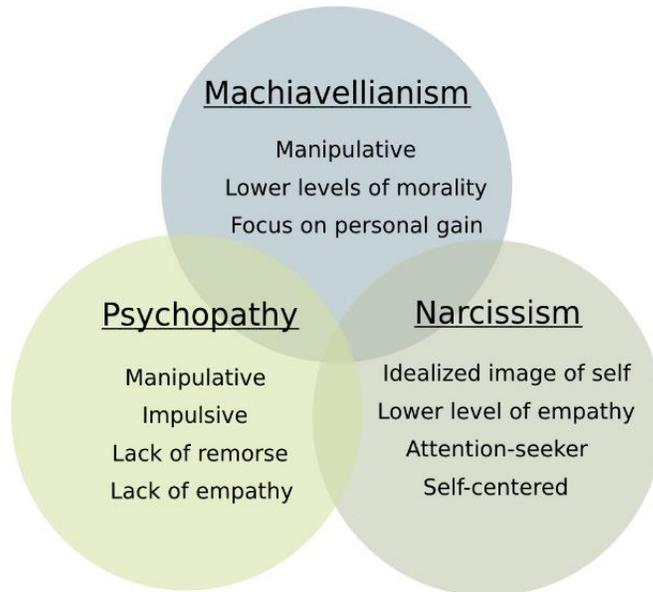

Figure 3. Dark Triad Personality Traits

## 3.2 Cognitive Processes

The mental processes through which an individual gains knowledge and comprehension are referred to as cognitive processes [117]. These processes constitute the human cognition which is a high-level functionality of the human brain and includes knowing, thinking, judging, remembering, and problem solving [117].

### 3.2.1 Information Processing

Information-processing refers to the way people handle and judge the credibility of information received from external stimuli [28]. There are several models that deal with the information-processing procedure in situations that external factors or actors attempt to persuade people. Such models are the Heuristic-Systematic Model (HSM), the Elaboration Likelihood Model (ELM), and the Social Judgment Theory (SJT) [26]. However, the use of the HSM has become more popular in phishing studies. Many cyber-security researchers consider the HSM as a framework that can assist in revealing the possible reasons that affect susceptibility to SE attacks [61,116,120].

As the HSM model describes, there are two primary ways that an individual may follow in information-processing: the *heuristic* way and the *systematic* way. The *heuristic* way is the less cognitive way, meaning that is shallower and requires less cognitive resources for information assessment. People following this way of thinking, rely less on the context and meaning of a received message and more on other factors, e.g., non-verbal factors [42]. On the other hand, the *systematic* way is a deeper way of information-processing and requires more cognitive effort, where people prefer to analyze information, and usually tend to make some research on their own to judge the credibility of received messages [57], [78].

---

* SD3 - https://openpsychometrics.org/tests/SD3/

### 3.2.2 Decision-Making

Information-processing is the first step towards decision-making: external stimuli produce some information, which is then processed, before making any judgments. Sometimes, people are called to reach a certain decision after information processing, e.g., proceed or not with an action. Bruce and Scott in [111], after reviewing the existing literature, pointed out that there are five distinct decision-making styles, namely: *rational*, *intuitive*, *dependent*, *avoidant*, and *spontaneous*. It is argued that usually, people present different levels of all styles in their decision-making process, although there is a dominant style for each individual [7].

The *rational* style of decision-making refers to an individual who relies on logical analysis and research towards reaching a decision, while on the other hand, the *intuitive* style implies a reliance on hunches, experience, and emotions. An individual who presents a higher level of the *dependent* style usually seeks support during the decision process and tends to consult others before reaching a final decision. Some individuals avoid decision-making as much as they can and, thus, they present a high level of the *avoidant* style. On the contrary, the *spontaneous* style attributes to impulsive, and rapid decision-making, where the individual wants to get through the process as quickly as possible [16].

The General – Decision – Making – Style Scale (GDMS) [111] can be utilized to evaluate the process of decision-making of an individual (the tendency to procrastinate or the *rational* style of decision processing, etc.). The relationship between personality traits, derived from the FFM, and decision-making styles has been examined and statistically significant associations have been discovered. It is worth mentioning that the personality trait of *neuroticism* has a positive correlation with the style of *spontaneous* decision-making, which may include high-risk towards security [16].

### 3.2.3 Risk-Taking

There are two types of risk-taking, namely *active*, and *passive* risk-taking. The *passive* type of risk-taking is defined as the decision to not proceed with an action that may be beneficial and is a strong predictor for investigating a user's intentions relating to cyber-security (e.g., not strengthening a weak password, etc.) [11]. *Active* risk-taking is related to high-risk actions that an individual might proceed to (e.g., downloading a file from an untrusted website, etc.) Active risk-taking can be measured by the DoSpeRT[*] (Domain-Specific Risk-Taking) Scale [123], regarding five sub-domains: *social, safety and health, ethical, financial, and recreational* [19]. On the other hand, passive risk-taking can be measured through the Passive Risk-Taking instrument (PRT) [71].

### 3.2.4 Protection Motivation

In [104] R. W. Rogers introduced the Protection Motivation Theory (PMT). According to PMT the possibility an individual may or may not proceed with actions related to one's protection is analogous to the following: the cognitive appraisal of external stimuli; the likelihood of noxious events; the level of belief that a certain mitigation approach may thwart the occurrence of the event in question. The PMT model comprises five factors attempting to measure and assess protection behavior, namely: *perceived severity, perceived vulnerability, self-efficacy, response efficacy, and response costs*. In [21,76,87] it is argued that all these factors are good indicators of the security behavior of users towards their electronic devices.

---

[*] DoSpeRT - https://www.midss.org/sites/default/files/domainspecificrisktaking.pdf

### 3.3 Security-Related Behavior Intentions

To identify a user's security-related behavior intentions, the Security Behavior Intentions Scale (SeBIS) has been developed by Egelman and Peer in [44]. The scale comprises 16 items that are grouped into four sub-domains, namely:

- *password generation* (e.g., strong passwords, use of password management tools, etc.)
- *system updates* (i.e., keeping software up to date)
- *device security* (i.e., locking devices)
- *proactive awareness* (i.e., considering security alerts and acting upon them)

The correlation between SeBIS and other psychometric instruments was examined by the authors in [44]. Results of their examination showcased a strong correlation between DoSpeRT and SeBIS. Moreover, both negative and positive correlation was observed between SeBIS and GDMS, regarding the tendency to procrastinate and the *rational* style of decision-making, respectively.

### 3.4 Persuasion Principles

According to R. Cialdini in [32], there are six distinct persuasion tactics, namely: *reciprocity*, *scarcity*, *commitment* (consistency), *social proof* (conformity), *authority*, and *liking*. Each one describes a different way in which individuals may be affected and influenced subconsciously [31,33]:

- The *reciprocity* principle is about people feeling obligated to return favors or help. That is, people are inclined to adhere to persuasive requests from people that they feel indebted to.
- When the *scarcity* principle is utilized as a persuasive method, keywords like "shortage", "limited availability", and "deadline" are often used. These words tend to pressure an individual to act quickly, out of fear of losing something valuable.
- The *commitment* (consistency) principle refers to one's promise to proceed with an action, living up to the expectations of others, or staying committed to/consistent with an idea. On top of that, people tend to feel even more obligated to complete an action if the commitment for this action was made publicly.
- The *social proof* (conformity) principle describes the tendency of humans to be influenced by the actions of other people, especially if they have personal or social characteristics in common with these people. A popular example of this principle is "canned laughter". This is a separate sound effect that imitates the sound of audience laughing and aims to provoke laughter to the viewer.
- *Liking* is a persuasion principle that has to do with the increased possibility of proceeding with an action if the latter is requested by someone one knows or likes.
- The *authority* principle is about people obeying trusted sources, whether this refers to proceeding with an action or to considering information coming from this source as more valuable.

Recently, a measure was developed to assess the level of susceptibility of an individual towards Cialdini's six principles of persuasion, the Susceptibility to Persuasion Scale (STPS) [85]. By utilizing this scale, researchers can measure the level of response in each one of the persuasive principles and therefore, this scale can serve as a good indicator of susceptibility to SE attacks, which usually employ several persuasion principles [53].

# 4 IDENTIFYING HUMAN VULNERABILITIES

In the context of HVA in cyber-security, it is mandatory to take into consideration several parameters in the endeavor to identify vulnerabilities. Personality traits, information-processing, decision-making, attitude towards security, level of risk-taking, and protection motivation, as well as the unique combination of these factors, play an important role in this venture. Additionally, there are several more factors that need to be taken into account, namely:

- demographic data of users (e.g., gender, age, culture)
- current public situations that attackers may use for their advantage (e.g., Covid-19),
- users' habits (e.g., online shopping, clicking on advertisements, etc.),
- users' profession (highly related to information technology or not, e.g., software engineers, teachers, etc.)
- familiarity with cyber-security issues [48].

## 4.1 SE Attacks and Attack Tactics

Another important variable in the cyber-security equation is the attacker's behavior and the attack tactic that is being followed. As mentioned before, the excessive increase in social media use has raised several concerns, since the human factor is considered a vulnerable point for exploitation. As the number of social media users increases, so do the number of vulnerable points that SE attackers can exploit. The authors of [107] and [126] proposed that SE attacks can be categorized as follows:

- *human-based* attacks, where the attacker interacts directly with the victim, and
- *computer-based* attacks, where the attacker interacts with many victims simultaneously by utilizing software.

According to [70,100,103] SE attacks can be grouped further into:

- *technical based*, where information gathering about the victim is achieved through the internet (e.g., compromised websites, etc.)
- *social based*, where the attacker develops a relationship with the victim and exploits the victim's trust and emotions.
- *Physical based*, where information gathering about the victim is achieved via physical actions (e.g., stalking, etc.)

Below, some examples of SE attacks are described in a few words, based on the taxonomy of SE attacks presented in [107]:

- *Impersonation*, where the attacker pretends to be a person of trust for attacking by telephone an organization's help desk [107]
- *Tailgating*, where the attacker enters a restricted area by following someone who has authorized access [126]
- *Pharming* through online platforms, where the traffic generated from a website is stolen and redirected to a fake website. There, any information that the victim enters can be accessed and stolen by the malicious actor [13]
- *Phishing*, which can be divided into several sub-categories (e.g., spear phishing, whaling phishing, etc.) and it can be conducted through phone, email, online platforms, etc., aiming to retrieve sensitive information from the victim via deception [102,107]
- *Shoulder surfing*, where the victim is being watched while entering sensitive information [107]
- *Ransomware*, where sensitive data have been encrypted by the attacker and the victim is blackmailed under the threat of data leakage. The payment of a ransom (usually in cryptocurrency) in a specific timeframe is demanded, for the victim either to regain access to the data or to prevent data publication [52,72].

All the different types of SE attacks and the sub-categories of phishing attacks are depicted in Figure 4, as presented in [107].

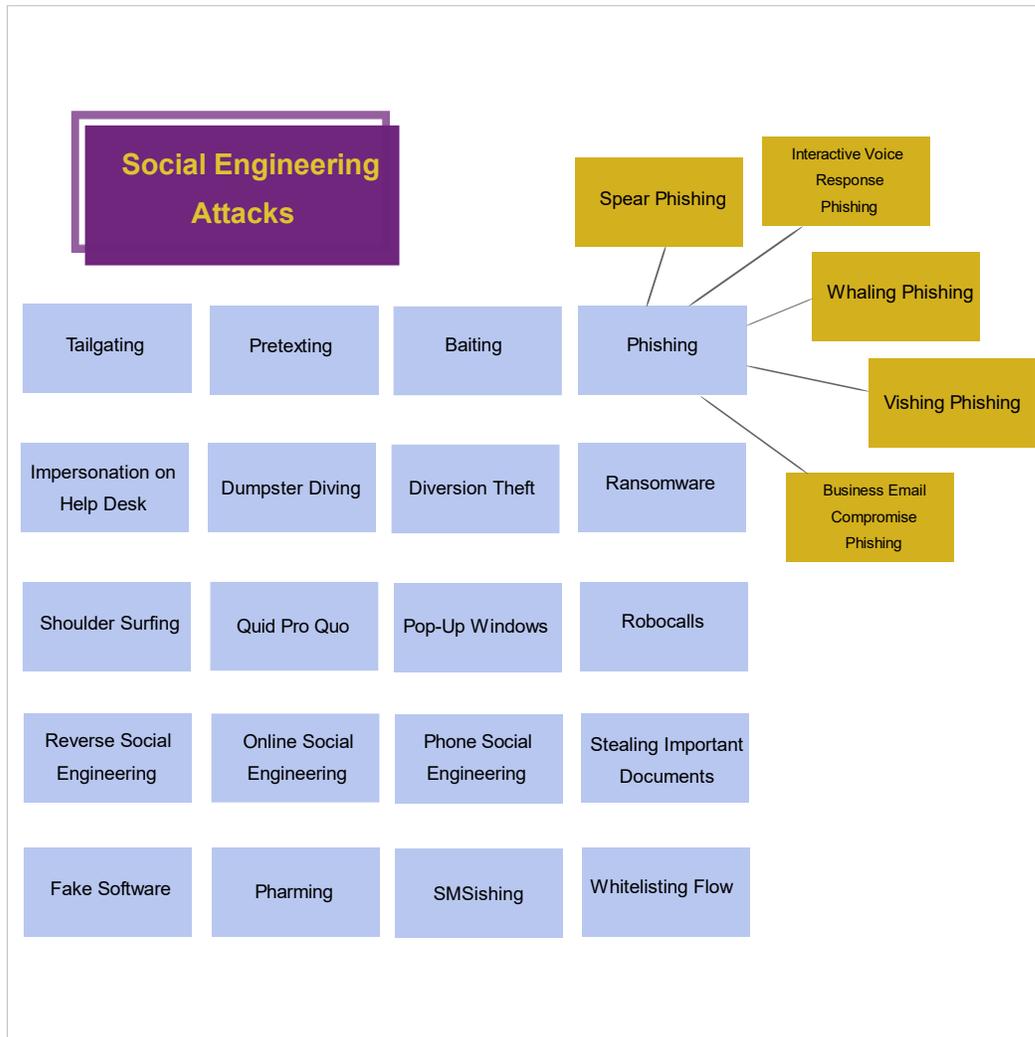

Figure 4. SE Attacks and Phishing Sub-Categories

However, SE attacks can be further categorized based on several other factors, such as execution location, or level of interaction with the victim. Ghafir *et al.* in [53] categorized SE into two subsets: *physical location* attacks and *psychological methods*. The first subset includes attacks like the aforementioned examples, i.e. impersonation, tailgating, phishing, pharming, etc. The second subset refers to attacks taking advantage of the six principles of persuasion, as well as the human tendency to provide help. What is more, in attacks targeting helpdesks of organizations, psychological methods usually take advantage of the low-level of involvement, as well as the possible high-level of ignorance of the victim, regarding the security policies and strategies of the organization. Moreover, attack patterns usually include four distinct steps [24]:

- research (information gathering about the target)
- hook (interaction with the victim)
- play (execution of the attack), and
- exit (interaction termination without leaving any traces).

Reasoning from this fact, SE attacks can be further grouped into two categories according to the level of interaction between the attacker and the victim, according to the authors of [24,99,108]:

- hunting (minimum interaction)
- farming (relationship establishment).

All aforementioned attack categories and sub-groups are gathered in Table II.

Sharma and Bashir, who focused their attention on attackers' footprints, explored attack tactics and specific language utilized in SE attacks, such as in phishing emails [112]. They conducted a content analysis on reported phishing emails that were provided by the Information Security Office of Berkeley University and the SecureIT-Kent State University. The analysis concerned date, time, content, subject, compelling words, and other information presented in phishing emails. Results revealed that phishing emails with a subject related to online accounts usually contain words such as *"suspend"*, *"update"*, *"declined"* and *"confirmation"*, with a tendency to stir up emotions like fear, trust, and anticipation.

Table II SE Attack Groups and their Sub-Categories

| SE attacks groups | Sub-categories | | | Paper |
|---|---|---|---|---|
| Information gathering | Human Based | | Computer Based | [107], [126] |
| Type of Interaction | Technical Based | Social Based | Physical Based | [70], [100], [103] |
| Level of Interaction | Hunting | | Farming | [24], [108], [99] |
| General Target | Physical Locations | | Human Vulnerabilities | [53] |

Moreover, it was observed that emails with a subject of payment/transaction or with a shared document aim to provoke curiosity, trust, and anticipation, for successfully launching the attack.

## 4.2 Human Vulnerabilities, Personality Traits, and Cognitive Processes

According to the degree that each personality trait is present in an individual's personality, one may be more susceptible to certain persuasion principles and less susceptible to others. The following subsections present the correlation between:

- human vulnerabilities and personality traits of the FFM
- human vulnerabilities and personality traits of the MBTI

- malicious traits and insider threats, which raise a human-factor risk
- human vulnerabilities and cognitive processes
- personality traits of the FFM, cognitive processes, and susceptibility to cyber-threats

### 4.2.1 Human Vulnerabilities and Personality Traits of the FFM

Relations between personality traits of the FFM and Cialdini's principles of persuasion were explored by Uebelacker and Quiel in [115], who proposed their own, theory-based, "Social Engineering Personality Framework". The authors assumed the following associations between certain persuasion principles and personality traits, based on their research of previous literature:

- *Extraversion* increases susceptibility to liking, scarcity, and social proof, as it is highly associated with sociability.
- Conscientiousness attributes to rules adherence, hence this trait increases susceptibility to authority, commitment/consistency (especially publicly made commitments), and reciprocity.
- High levels of the *agreeableness* trait increase vulnerability towards SE attacks in general, and especially towards persuasion principles like social proof, authority, reciprocity, and liking, as people with this trait tend to be more trusting of others.
- The *openness* trait also implies increased vulnerability towards SE tactics; however, research has revealed that this trait is also highly correlated to computer proficiency, which leads to a decrease in susceptibility. Thereby, the authors assumed that openness may be related only to the scarcity principle, out of fear of losing something valuable.
- *Neuroticism* is associated with computer anxiety and increased security-related reservations as mentioned before, which decrease susceptibility; however, people scoring high on this trait may be vulnerable towards the authority principle, according to the authors, due to their tendency to comply with instructions and commands from authorities.

Nevertheless, these assumptions require further validation, while the proposed framework remains a theoretical approach yet to be implemented.

### 4.2.2 Human Vulnerabilities and Personality Traits of the MBTI

A methodology for evaluation of human vulnerabilities was proposed by Cullen and Armitage in [39], based on the personality traits proposed by the MBTI. During their research, phishing emails were sent to the subjects, each one containing a different persuasion principle. Each participant represented a specific personality trait of the MBTI, and all participants were asked to select which email was more appealing to them. Results of this research showcased the following:

- the *extrovert* profile was more susceptible to the *liking* principle
- the *sensing* profile was more susceptible to *commitment*
- the *thinking-feeling* profiles were more susceptible to *authority* and *social proof*, respectively
- the *judging-perceiving* profiles exhibited greater susceptibility to *reciprocation* and *distraction*, respectively.

### 4.2.3 Human Vulnerabilities and Malicious Traits

Recent studies are starting to focus their interest on the connection of malicious activity to human behavior and personality traits, in an attempt to understand cyber-attacks' tactics and predict cyber-security incidents. Studying and measuring maliciousness though, offers not only comprehension of attackers' behavior and an advantage in the development of defense approaches, but also assists in predicting insider threats, which are considered human-factor vulnerabilities [124].

Several tools of measuring "dark" personality traits and personal characteristics are starting to be utilized in the context of measuring maliciousness. For example, Curtis *et al.* in [40] investigated the link between personality traits of the Dark Triad, phishing effort of the attacker, and attack's success, alongside user susceptibility to phishing emails. Results showcased that the participants of the attackers' group with the personality trait of *machiavellianism* put the most effort into the composition of phishing emails. On the contrary, participants with the personality trait of narcissism put less effort, maybe due to their overconfidence, according to the authors. It is worth mentioning that when emails were composed by attacker-participants with the trait of *narcissism* and then presented to victim-participants with the same trait, an increase in susceptibility was observed. Moreover, narcissistic end-users appeared to be the most vulnerable group to phishing emails. However, the authors mentioned that Dark Triad personality traits of attackers are not enough to predict the success of a phishing email.

The issue of maliciousness was also addressed in the study of King *et al.* in [73], who thoroughly reviewed the related literature of other research domains like sociology, law, psychology, etc. According to the authors, several factors may be indicative of maliciousness such as mental stability, personality traits, self-perception, biases and attitudes, emotions, norms, government structure, etc. All these indicators are grouped into four categories in their study, namely:

- *individual level* (personality traits and mental process)
- *micro-level* (interpersonal interactions)
- *meso-level* (subcultures and group membership)
- *macro-level* (how cultures – large or national – affect maliciousness).

Instruments like the Dark Triad questionnaire in combination with the BFI, and the use of the Levenson Self-Report Psychopathy Scale (LSRP)[*], which measures mental stability [75], are proposed to be utilized in terms of assessment at the individual level. Metrics like Hofstede's cultural dimensions [63] can be used for the assessment of cultural values like patriotism, censorship, civil disobedience, and military success, characteristics which have been found indicative of the level of maliciousness, regarding assessment at the macro-level.

### 4.2.4 Human Vulnerabilities and Cognitive Processes

The aim of Goel *et al.* in [80] was to study user susceptibility to deception in relation to information-processing. The authors proposed a theoretical framework to investigate how certain factors like the context of phishing emails may affect susceptibility. In this study, 7225 undergraduate students were recruited. Each participant was presented with one of eight different types of phishing emails containing a link. Half of the emails attempted to provoke fear of losing something (e.g., tuition assistance), while the other half sought to induce anticipation of acquiring eligible stuff (e.g., a mobile phone). Results of this study revealed that the possibility of losing or gaining something is a factor that plays a critical role in susceptibility to phishing. Regarding context of phishing emails, those that specifically targeted issues and desires of the victim were found to be more deceptive. Furthermore, the study revealed that 68.7% of the students that opened the emails also clicked on the link. Women exhibited a higher rate than men in opening the emails but the same rate in clicking the link. Students of the business major were also more likely to open the emails, but they had the same rate of clicking as students of the humanity major. The authors suggested that these results indicate that individuals with high levels of curiosity or with job/education roles that demand higher rates of human interaction, through emails, are more likely to open phishing emails; however, falling victims of the deception techniques utilized is not always the case for these individuals.

---
[*] LSRP Scale - https://openpsychometrics.org/tests/LSRP.php

Similarly, Yan *et al.* in [127], studied the way a subset of ordinary users (i.e., individuals who are familiar with cyber technologies but have not received any cyber-security training), processes information (*heuristic* vs. *systematic*) in light of cyber-security incidents. The authors also examined the way these users assess cyber-security issues, in an endeavor to identify the weakest link among weak-link users. A sample of 462 undergraduate students of different gender, age, cultural backgrounds, and majors was chosen. During this research, the Cyber-security Judgement Questionnaire was developed, comprising 16 real-life cyber-security scenarios – 8 risky actions and 8 safe cyber-security actions. The participants of the study were asked to judge the scenarios as risky or safe. The authors observed that 65% of all students judged the scenarios correctly, while on the other hand, 23% of the students misjudged more than half of the scenarios. It should be noted that no significant difference between *heuristic* and *systematic* information-processing was observed. According to the authors, this may be the case due to the lack of knowledge that students have concerning cyber-security, which places a limit on their *systematic/rational* information-processing and judgment regarding cyber-security issues. An observation like this suggests that *heuristic* and *systematic* information-processing assessment, in cases of cyber-security incidents, may provide more insight when it is focused on users that already have a level of familiarity with cyber-security.

*4.2.5 Personality Traits, Cognitive Processes and Susceptibility*

For the last few years, researchers have been trying to unravel the relations between personality traits, cognitive process, and susceptibility to cyber-threats. In this context, Cho *et al.* in [65] proposed a mathematical model using Stochastic Petri Nets to investigate how specific personality traits of the FFM can affect the decision-making process and the level of trust and risk a person showcases when encountering possible cyber-threats. Both genuine and phishing emails were presented to the participants of the study, who were asked to judge the presented information as trustful or distrustful. The results of their research showed that personality traits like *neuroticism* and *agreeableness* play an important role in the process of decision-making. Under circumstances of great uncertainty about the decision, the aforementioned traits may influence the degree in which the person will trust or not the received information. Another finding worth mentioning is that *neuroticism*, combined with low *openness* and low *conscientiousness*, leads to a decrease in decision performance, which is the ability to detect cyber-threats. As a result, this may induce an increase in susceptibility.

The link between FFM and the HSM, concerning susceptibility to phishing in social networks, was investigated by Frauenstein and Flowerday in [48]. In this research a sample of 215 final-year, undergraduate students, actively engaged in social networks, was studied. Results suggested that *heuristic* information-processing is highly associated with susceptibility to phishing in social networks. Additionally, it was observed that participants with the personality trait of *conscientiousness* were found to be the least vulnerable group to phishing. The authors argued that this may be the case due to the negative correlation between *conscientiousness* and the *heuristic* information-processing style. That is, individuals with the *conscientiousness* trait tend to follow a more *systematic*, thus rational, way of information-processing.

**4.3 Human Vulnerabilities, Demographics, and other Personal Characteristics**

Personal characteristics like gender, age, profession, cultural background, computer experience, etc., may influence the way an individual assesses cyber-security incidents and issues. The following sub-sections refer to studies that elaborate the effect of age and gender in susceptibility to cyber-attacks, as well as to studies that attempt to explain the relations between susceptibility and other personal characteristics (e.g., impulsivity, susceptibility to certain persuasion principles, computer experience and proficiency).

### 4.3.1 Gender and Age

Orji *et al.* in [95] aimed to identify whether susceptibility to Cialdini's persuasion principles is influenced by age or gender. The STPS [85] was utilized and data from 1.108 participants were analyzed. The participants of this study were grouped regarding gender, into males and females. The same participants were also grouped into adults (over 35 years) and young adults (18-25 years) regardless of gender. Results showcased that women were more susceptible to persuasion principles in general than men. Several differences were also observed regarding the degree in which adults and young adults were persuaded. It was revealed that *commitment* and *reciprocity* are the most persuasive principles, with *consistency* and *scarcity* ranking last, regardless of age and gender group. Females were found to be more susceptible to *reciprocity* and *commitment* than males. Adults were also more responsive to *commitment* than younger adults. Lastly, *scarcity* was found to be one of the least persuasive principles; however, the male and the young adult groups were more susceptible to it than the female and adult groups, respectively.

The gender factor and its effect on cyber-security behavior was also explored by Anwar *et al.* in [10]. The authors investigated how security is perceived and how security behavior is shaped with regard to gender. Utilizing the Health Belief Model by I. M. Rosenstock [106] and the PMT [104], they conducted an online survey involving 481 participants with job descriptions related to technology, either part-time or full-time. The implemented questionnaire involved the following categories of questions: "*self-reported cyber-security*", "*security self-efficacy*", "*peer behavior*", "*perceived severity*", "*cues to action*", "*perceived vulnerability*", "*response efficacy*", "*perceived benefits*", "*perceived barriers*", "*computer skills*", "*prior experience with cyber-security*" and "*internet skills*". Their research revealed that there are some major gender-related differences about how males and females self-report about their security behavior; however, authors suggested that this may be the result of the overconfidence of men and does not necessarily mean that women are more vulnerable to cyber-threats. Apart from that, results showcased statistically significant differences regarding *computer skills, cues-to-action, prior experience,* and *security-related self-efficacy* between male and female employees.

Similarly, McGill and Thompson in [82] utilized the factors of the PMT model in an anonymous questionnaire. In contrast to the research conducted by Anwar *et al.*, two extra factors were included in the questionnaire of this work, namely: *descriptive norm* (the belief that others possibly implement security measures) and *subjective norm* (the belief that others may wish for an individual to adopt security behavior). Data were collected from 624 participants and results illustrated that *perceived severity* was higher among females; however, females did not exhibit higher *perceived vulnerability*. Moreover, contradictory findings regarding females were also observed in the context of *descriptive* and *subjective norm*. Females scored higher than males regarding the *descriptive norm*. However, there was no significant difference, among gender, regarding the *subjective norm*. Some similar findings to the research of Anwar *et al.* in [10] were also observed. Males were found more likely to have attended security training in the past. Males also evaluated their security skills at a higher level than females. It is worth mentioning, that both described studies reached one similar conclusion, which referred to the importance of developing personalized and gender-specific security solutions and training programs.

As mentioned above, demographic factors like age and gender have been studied by researchers of the field. However, older age groups (60-90 years) were neglected from research until recently [43]. These days, older adults are more involved with technology and many of them are getting comfortable with social networks [113]. In this context, Ebner *et al.* in [43] developed the Phishing Internet Task (PHIT) framework to examine how susceptibility to online threats is shaped regarding older age groups. They recruited 157 web-active participants who were categorized into three groups: young adults (18-37 years), young-old adults (62-74 years), and middle-old adults (75-89 years). Daily mood of the participants was recorded by utilizing the PANAS questionnaire, which assess positive and negative mood [122]. Daily cognitive status of the participants was recorded too. Results of the study revealed that higher short-term memory and specifically higher short-term episodic memory in middle-old users decreased susceptibility to spear-phishing when associated with verbal fluency and high levels of positive mood. Furthermore, positive mood was also found to be positively correlated with increased susceptibility awareness towards phishing.

*4.3.2 Other Personal Characteristics*

The susceptibility to Cialdini's persuasion principles and its correlation to susceptibility to phishing attacks was researched by Parsons *et al.* in [98]. The authors tried to determine whether the STPS can assist in the prediction of users' susceptibility to phishing attacks. They conducted a web-based survey where they sent both genuine and phishing emails to 985 Australian participants – whose work involved at least some time using a computer. They used seven types of emails, one where no social influence principle was used, and six emails where each one was designed using one of Cialdini's six persuasion principles. Results revealed that emails based on the *reciprocity* principle were recognized as legitimate, most of the times, even when they were not. On the other hand, emails based on the *scarcity* principle were more possible to be recognized as frauds, even when they were genuine. The authors argued that this may be explained by the extensive use of this principle in real-life phishing emails, which has caused some kind of immunity to the users (i.e., they easily recognize this persuasion technique). It is worth mentioning, that most of the participants were unable to identify a legitimate URL from a phishing one. Moreover, participants who were found specifically susceptible to a certain persuasion principle, through the STPS, showcased an increased susceptibility to phishing emails that were authored based on the same principle. This observation suggests that the STPS can assist in vulnerability assessment and provides evidence that personalized training upon susceptibility could be a very useful tool to enhance cyber-security awareness. What is more, demographic data like gender, as well as other personal characteristics like impulsivity, time spent in front of a computer, and susceptibility to *social proof* were found to play an important role in the users' ability to detect deception and identify phishing emails.

From another point of view, computer experience and proficiency are also considered to be personal characteristics that play an important role in susceptibility to cyber-attacks., In this context, Ovelgönne *et al.* in [96] observed and analyzed data for more than 8 months by utilizing 1.6 million machines of professionals, software developers, and gamers, in an endeavor to unravel this relationship. For the scope of their research, the authors used the Symantec's WINE datasets [41] - comprised of real-world antivirus-software data – to study the behavior of users. The study was conducted by measuring and analyzing the number of applications and executable files detected in the users' machines. The travel history of the user (to physical locations) was also studied for determining the user's tendency to choose connectivity over security by connecting to the internet through free WIFI networks, which may be untrusted (e.g., through the free WIFI network of the Municipality of Athens during an afternoon walk downtown). It was observed that 83% more malware attacks occurred in gamers' rather than in non-gamers' machines. What is more, 33% more attacks occurred in machines of professionals than in non-professional users, suggesting that computer experience and expertise need to be considered as risk factors, regarding cyber-security. It was also observed that the software developers' machines contained a significantly higher amount of malware in contrast to other machines. The findings of this study indicated a relationship between the amount of malware on a machine and the amount of downloaded, unsigned, low-prevalence (not commonly detected in the dataset) binaries among all groups of users. It is noteworthy that the group which showcased the highest cyber-security risk was the software developers.

### 4.4 Stress, Anxiety, Time Pressure, and Human Vulnerabilities

According to the Yerkes-Dodson Law (YDL) [128], moderate levels of alertness can be beneficial and may serve as motivators, while low or high levels of alertness may result in poor performance. For many years, work managers – based on YDL – used to adopt strategies of manipulating stress levels of employees for boosting performance and increasing productivity [80]. However, recent studies in several scientific fields showcase contradictory results regarding stress levels and productivity. As mentioned earlier in this section, the personality trait of neuroticism is highly associated with increased levels of anxiety and alertness, hence it is suggested that individuals scoring high at this trait may be less susceptible to SE attacks [115].

In a similar manner, *stress* and *difficulties* experienced when trying to counteract an ongoing phishing attack have been found to promote users to be more aware of phishing, hence making them more capable of detecting a future attack. Chen *et al.* in [27] conducted a survey to explore how phishing susceptibility is affected by prior phishing encounters. They also aimed to identify the role of detection difficulty and detection failure, in the overall susceptibility of users. The authors recruited a sample of undergraduate students from a large public university and asked them to report a recent experience of facing a phishing attack. Results of this survey revealed that difficulties in the detection process, as well as the negative outcome of an experienced attack, have a significant effect on user susceptibility. Users who failed to detect a previous phishing attack were found to detect deception easier. According to the authors, this may be the case due to knowledge gained by the negative outcome and failure of the past, which led to increased alertness. Moreover, users with *detection self-efficacy* (i.e., users who have successfully detected a phishing attack in the past) have a greater possibility to detect future attacks. However, when users become overconfident about their detection abilities, they may be less focused and thus more susceptible.

From another point of view, non-secure human behavior can be shaped not only by personality traits and other personal characteristics but also by current circumstances that may cause anxiety to the user, such as *time-pressure* in the work environment. Chowdhury *et al.* in [29] developed an interview study and based on its results they managed to construct an integrative framework able to conceptualize *time-pressure* in the work environment. They also focused on the solutions that can be adopted, for the negative consequences of *time-pressure* to be decreased. They used a pool of participants consisting of non-security professionals, cyber-security experts, and private users (i.e., home users and students) whom they interviewed. After the interviews they were able to identify six distinct patterns of non-secure human cyber-security behavior, namely: *avoiding, bypassing, disclosing, disregarding, influencing,* and *over-relying*. They also observed that different moderating factors exist, regarding how *time-pressure* affects cyber-security behavior: *workplace characteristics, user characteristics,* and *task characteristics*. Finally, they suggested four layers of countermeasures that could be adopted to deal with cyber-security threats due to time pressure:

- operational countermeasures (training and awareness programs, security policies and procedures, data backups, simulation of security incidents)
- human countermeasures (security culture, behavior reinforcement, work life balance, and work design),
- technical countermeasures (restrictions, analytics and automation, security notifications, security tools)
- physical countermeasures (device labelling, and visual cues at workplaces).

### 4.5 Proposed User Susceptibility Profile

According to the information presented in Sections 3 and 4, a proposed user susceptibility profile is depicted in Table III, where all the individual characteristics that were found to affect human susceptibility to cyber-threats are gathered alongside the models or scales that are used for their assessment. Moreover, according to the literature that was reviewed in Section 4, the associations or correlations that were found between individual characteristics (e.g., personality traits, cognitive processes, etc.) and the level of susceptibility, either general (high/low) or specific (e.g., persuasion principles) are described in Table IV.

### 4.6 HVA Frameworks

With so many factors affecting and being affected by human vulnerabilities, it is crucial that certain mitigation approaches are developed, including educational and awareness-raising training regarding cyber-security. What is more, awareness-raising training needs to be personalized, as every person is susceptible to different cyber-threats and presents a different level of computer knowledge and skills. Hence, it is evident that HVA frameworks are necessary for assessing the cyber-security capacity of organizations and industries, as well as for identifying individual weaknesses, eliminating weak points, and strengthening cyber-defense strategies. Widdowson and Goodliff in [55] introduced the Cyber Human Error Assessment Tool (CHEAT) a model that was developed to provide an evaluation of an organization's cyber-security status with a specific focus on human vulnerability. In this tool, human factors, such as job-satisfaction, lack of power, password memory, workarounds, incident recording, and monitoring, productivity versus cyber-security, incidents and "near misses", and many others are mapped into five groups. These groups are *people, organization, technology, environment*, and *history*. CHEAT can be utilized for cyber-security vulnerability assessment regarding human factors, as well as for improving system design, training procedures, and risk-management decisions.

Furthermore, the analysis of users' behavior in social media can provide insights on users' susceptibility and assist in vulnerability identification. This is what the model, proposed by Abubaker and Boluk in [1], aims to achieve. The framework comprises 17 scenarios that cover 3 sub-groups of cyber-attacks, namely: *modern*, *classical attacks*, and *attacks targeting children*. Users need to answer several questions (e.g., duration of online time) for their online behavior and vulnerabilities to be assessed. The level of vulnerability for each user is predicted with a high accuracy rate, while the model can shed light upon cultural impacts on user susceptibility to cyber-threats.

Another framework that gathers and assesses information from social media use is the one proposed by the DOGANA project. DOGANA was introduced by Pacheco and Escravana in [97] with the goal to provide a framework that will be able to conduct Social Driven Vulnerability Assessment (SDVA). The model also takes into consideration ethical, psychological, and privacy features. The project consists of four phases for conducting SDVA: firstly, information about the participants needs to be gathered (by the organization, through social media, etc.); secondly, the approach of the attack needs to be prepared; the third phase refers to the execution of the attack; lastly, results of the attack analysis are gathered and presented as a vulnerability assessment report. Through all these phases, privacy is maintained and sensitive data, which can be used for identification of the participants, are discarded. As part of the DOGANA project,

Table III Proposed User Susceptibility Profile

| MODEL | \multicolumn{6}{c}{USER PROFILE — Factors & Sub-domains} | Paper |
|---|---|---|---|---|---|---|---|
| \multicolumn{8}{c}{**PERSONALITY**} |
| \multicolumn{8}{c}{**Personality Traits**} |
| **MBTI** | Sensing / Intuition | | Extraversion / Introversion | | Judging / Perceiving | | Thinking / Feeling | [88] |
| **Dirty Dozen / SD3** | Psychopathy | | Machiavellianism | | | | Narcissism | [67], [101] |
| \multicolumn{8}{c}{**COGNITIVE PROCESS**} |
| \multicolumn{8}{c}{**Information Processing**} |
| **HSM** | Heuristic Approach | | | | Systematic Approach | | | [28] |
| \multicolumn{8}{c}{**Decision Making**} |
| **GDMS** | Avoidant | Dependent | Spontaneous | | Rational | | Intuitive | [111] |
| \multicolumn{8}{c}{**Protection Motivation**} |
| **PMT** | Perceived Severity | Perceived Vulnerability | Self-efficacy | | Response Costs | | Response Efficacy | [104] |
| \multicolumn{8}{c}{**Risk-Taking**} |
| \multicolumn{8}{c}{*active risk*} |
| **DoSpeRT** | Social | Safety & Health | Ethical | | Financial | | Recreational | [123], [19] |
| \multicolumn{8}{c}{*passive risk*} |
| **PRT** | Resources | | Medical | | | | Ethical | [71] |
| \multicolumn{8}{c}{**MOOD**} |
| \multicolumn{8}{c}{**Postive affect**} |
| **PANAS** | Alert | Strong | Determined | | Enthusiastic | | Proud | [122] |
| | Active | Inspired | Attentive | | Excited | | Interested | |
| \multicolumn{8}{c}{**Negative affect**} |
| | Upset | Distressed | Scared | | Irritable | | Afraid | |
| | Jittery | Ashamed | Guilty | | Nervous | | Hostile | |
| \multicolumn{8}{c}{**BEHAVIOR**} |
| \multicolumn{8}{c}{**Security-Related Intentions**} |
| **SeBIS** | Password Generation | System Updates | Device Security | | | | Proactive Awareness | [44] |
| \multicolumn{8}{c}{**Susceptibility to Persuasion**} |
| **STPS** | Authority | Scarcity | Liking | | Commitment | Reciprocity | Social Proof | [85] |
| \multicolumn{8}{c}{**Maliciouss Activity**} |
| **LSRP** | Primary Psychopathy (emotional affect) | | | | Secondary Psychopathy (lifestyle) | | | [75] |
| **Hofstede's cultural dimensions** | Patriotism | Censorship | Civil Disobedience | | | | Military Success | [63] |
| \multicolumn{8}{c}{**OTHER DOMAINS**} |
| | Demographic Data (e.g., gender, age, cultural background, etc) | Profession (highly related to information technology or not) | Familiarity with Cybersecurity Issues | | Stress, Anxiety, Time Pressure in the work environment | Current Social Situations | Habits | |

Frumento *et al.* in [49] proposed a six-layer theoretical model, to be utilized for the second phase of the SDVA, that would be responsible for creating victim templates to discover the most possible attack path (HAV - Human Attack Vector) that can be used for targeting a particular individual. Their model is called Victim Communication Stack (VCS) and consists of the following layers in exact order, each one connected to its predecessor: *persona modeling, semantic, syntax, medium, device*, and *context. Persona modeling* relates to personality traits, demographics like gender and age, social role or job-related role, and cultural background, all found to play a critical role in victim susceptibility to SE attacks. The *semantic* layer includes the speculations of the attacker about the victim's mindset. The *medium* layer refers to the medium used to launch the attack. The *device* layer concerns the possible device through which the targeted individual might experience the attack. Finally, the *context* layer attributes to environmental characteristics, such as time and place of the attack.

Another user-centric framework that gathers information from social media use is proposed in [2] by Albladi and Weir. This framework gathers all user characteristics that affect decision-making when one encounters a cyber-security threat in social networks. It is based on four principles:

- *socio-emotional* (fear and anxiety tactics, trust tactics)
- *habitual* (network engagement, number of friendship connections, time spent online)
- *perceptual* (perceived vulnerability, perceived severity, response-efficacy, self-efficacy) and
- *socio psychological* (Big Five Personality traits, demographic features, computer experience, and education).

The framework has been evaluated and validated by information security experts; however, further research is required for its automation.

A situation of tampered or breached data can be a major issue for most organizations and industries. Hence, human susceptibility to cyber-threats may result in affecting an organization's public image towards its clients which may lead to damages, financial or not. During their research upon human vulnerabilities, Ani *et al.* [9] observed that many employees acquired the necessary knowledge to identify cyber-security attacks or incidents. However, not many of these employees had the skills to counteract against the detected issues. Hence, the authors proposed a framework that can quantify the cyber-security knowledge and skills of employees, while it can also identify weak-link users, to provide knowledge about specific issues that need to be addressed for enhancing the security of a particular industry or organization. The proposed framework comprises five levels of evaluation, namely: *definition* (of skills, knowledge, security capabilities, and baselines), *data collection* (information extraction about entities), *formulation* (mathematical procedures to compute capability values), *representation* (capacity visualization), and *attribution* (identification of the weakest link).

On another note, Hughes-Lartey *et al.* [64] proposed a hybrid framework comprising human factors and technological aspects, for assessing vulnerabilities and minimizing attacks on data locations. The human factor of the proposed framework assesses four subdomains, namely:

- *management* (security policies, support, and prioritization from senior departments, effective communication of security risks)
- *environment* (Information security culture, national culture, regulations, and standards)
- *preparedness* (training, awareness, change).
- *responsibility* (monitoring and control of security policies, reward, and deterrence for promoting security behavior, acceptance of responsibility by all participants).

In Table V the aforementioned frameworks and the authors that proposed them are showcased.

For the development of HVA frameworks to be possible, a taxonomy of human vulnerabilities is needed. One of the most recent studies in the context of SE and human vulnerabilities was conducted by Wang *et al.* in [121]. The authors proposed a conceptual model to assist in the comprehension of the way SE attacks work and take effect. They managed to identify more than 30 effect mechanisms, including persuasion principles, trust, deception, and distraction techniques, and cognitive approaches, which may be utilized by attackers to exploit human vulnerabilities. The proposed conceptual model included more than 40 human vulnerabilities that appear to play a crucial role in susceptibility to cyber-threats. The included vulnerabilities were grouped into four distinct categories, namely:

- *cognition and knowledge*
- *behavior and habit*
- *emotion and feeling*
- *psychological vulnerabilities*, which were further grouped into human nature vulnerabilities, and vulnerabilities deriving from personality traits or one's character.

This conceptual model does not serve as an HVA framework; however, the taxonomy-like map of human vulnerabilities that was proposed may provide insight into SE attack methods and approaches, as well as into the mechanisms that shape human susceptibility, and the way users respond to these attacks. Moreover, the aforementioned work may assist in the development of future HVA frameworks or in the enhancement of already existing frameworks.

## 5 EDUCATION AND AWARENESS

Poor security strategies may result in exposing a firm to several cyber-threats and making it a potential target for attackers, like in the case of the Anonymous attack against HBGary Federal in 2011 [74]. Weak password management, inferior software patching, and lack of proper distribution of sensitive data were some of the exploited loopholes that led HBGary to fall prey to malicious actors [59]. Hence, it is evident that regular risk assessment, continuous awareness-raising training of employees, as well as constant vulnerabilities testing and defense strategies are of paramount importance for ensuring cyber-security in an organization, as it has already been stated by early studies [59].

### 5.1 Defense Approaches and Mitigation Strategies

Some of the current defense approaches include the improvement of physical security and the enhancement of security policies, as well as continuous awareness-raising and incidence-handling training programs. Ghafir *et al.* in [53] suggested that all these approaches and techniques should be adopted into a multi-layer security shield, rather than be single approaches, to mitigate possible risks. Similarly, Conteh and Schmick in [35] recommended several measures that a "Defense in Depth" structure (multi-layer defense) should include to be effective against SE attacks. The proposed strategy includes the following: *security policy, education and training, network guidance, audits and compliance, technical procedures*, and *physical guidance*.

Evans *et al.* in [47] proposed a framework that is able to identify and counteract against human error, by integrating the following:

- a Human Reliability Assessment instrument (HRA), which includes quantitative and qualitative techniques to evaluate risk regarding human error
- a Statistical Quality Control method (SCQ) commonly used in manufacturing environments
- a vulnerability scoring system that focuses on human rather than technical vulnerabilities

Table IV Factors Affecting Cyber-Security Risk

| Factor | | | CYBERSECURITY RISK | | | | | | | | | Paper |
|---|---|---|---|---|---|---|---|---|---|---|---|---|
| | | | High Susceptibility | | | | | | | Low Susceptibility | | |
| | | | General | Specific | | | | | | General | Specific | |
| Domain | Model | Characteristic | | Liking | Scarcity | Commitment | Authority | Social Proof | Reciprocity | | Phishing | |
| PERSONALITY TRAITS | FFM | Extraversion (high) | | × | × | | | × | | | | [115] |
| | | Agreeableness (high) | × | × | | | × | × | × | | | [115] |
| | | Neuroticism | | | | × | | | | | | [115] |
| | | | × | | | | | | | | | [65] |
| | | Openness | | | | × | | | | | | [115] |
| | | Conscientiousness (high) | | | | × | × | | × | | | [115] |
| | | | | | | | | | | | × | [48] |
| | MBTI | Extraversion | × | × | | | | | | × | | [39] |
| | | Thinking | | | | | × | | | | | [39] |
| | | Feeling | | | | | | × | | | | [39] |
| | | Judging | | | | | | | × | | | [39] |
| | | Perceiving | × | | | | | | | | | [39] |
| | | Sensing | | | | × | | | | | | [39] |
| | Dark Triad | Narcissism | × | | | | | | | | | [40] |
| COGNITIVE PROCESS | HSM | Heuristic Approach | × | | | | | | | | | [48] |
| | | Systematic Approach | | | | | | | | × | | [48] |
| DEMOGRAPHICS | - | Gender: Males | | | | × | | | | | | [95] |
| | | Gender: Females | | | | | × | | | | × | [95] |
| | | Age: Young Adults (18-25) | | | | × | | | | | | [95] |
| | | Age: Adults (over 35) | | | | | × | | | | | [95] |
| OTHER PERSONAL CHARACTERISTICS | - | Computer expertise | × | | | | | | | | | [96] |
| | | Profession: Software developers | × | | | | | | | | | [96] |
| | | Previous Victimization | | | | | | | | × | | [27] |
| | | Previous Self-efficacy in attack detection | | | | | | | | × | | [27] |
| | | Overconfidence | × | | | | | | | | | [27] |

Among various available methods for HRA, the authors suggested the use of the Human Error Assessment and Reduction Technique (HEART). According to HEART, tasks are categorized into nine generic types including *complex, simple, heterogeneous,* and *routine* task types, *familiarity/unfamiliarity* of the user, *correct response* to incidents, and *system restoration* with or without following specific procedures. The framework is proposed to be run in all stages of a Plan-Do-Study-Act cycle (PDSA) for achieving full awareness and accomplishing enhancement of effectiveness in human error handling.

Priming and warning are often used for informing, raising awareness, and alerting people about potential threats. Priming refers to the condition when an individual's behavior can be influenced and changed due to past exposure to several types of stimuli, such as visual patterns, objects, sounds, or words [110]. The use of priming in raising awareness about threats aims to provoke suspicion [69]. Warning refers to certain messages and signs that tend to inform people of danger [69]. Junger *et al.* in [69] investigated the effectiveness of priming and warning in raising awareness and preventing people from disclosing sensitive, personal information. The authors studied 256 visitors of a shopping district in a town in the Netherlands by handing them questionnaires. Participants were segregated into three distinct groups, namely: a group that had to answer 4 questions about cyber-attacks prior to the questionnaire (priming condition); a group that was handed a warning leaflet before filling in the questionnaire (warning condition); a control group where the participants answered the questionnaire without being primed or warned beforehand. Results showcased that the interventions failed to prevent the participants from disclosing information, with 80% of the total sample disclosing information in general, while 43.5% gave away information about their bank accounts. Furthermore, another unexpected observation was made: the warning condition appeared to have a negative effect on the disclosure of information. In the case of requesting details about online shopping, 98.7% of the participants filled in the name of the online shop they had recently visited, in comparison to 87.4% and 89.1% of the control and priming group, respectively.

Priming and warning, as described above, rely on establishing a recent past experience – often referred to as implicit memory [110] – for altering future behavior of an individual. In the context of cyber-security, automated cyber-attack detectors may protect users from encountering many cyber-security issues. That is, detectors may assist individuals in handling cyber-security issues, or they may address incidents without user interaction. However, this impedes users from gaining experience in cyber-security. Along these lines, Sawyer and Hancock in [109] addressed the existence of the prevalence paradox in cyber-security and its success rate, when used as a type of cyber-attack. The prevalence paradox describes that when an event occurs in high frequency, it is more possible to be detected. On the contrary, when an event occurs seldomly then the possibility to be detected decreases logarithmically. The authors recruited 30 undergraduate students who, at first, attended a brief training on phishing emails (both legitimate and malicious emails were utilized as a demonstration). After the training, the participants were presented with 300 emails, and they were given the following choices of action: download email attachments; upload their own attachments and reply to emails; report the email as suspicious. The subjects of this study were informed that they were taking part in a phishing detection study; however, they were ignorant of the rate of genuine and malicious emails that they encountered. The authors wanted to observe whether the prevalence paradox would appear in low rates of suspicious emails - 1%, 5%, and 20% occurrence probabilities were used. Attacks that were presented at an occurrence probability of 1% resulted in a lower accuracy rate of detection by the participants. These results showcased that while automation in cyber-attack detection rises, the ability of users to efficiently detect and report malicious actions decreases. Raising awareness or developing enhanced security systems alone is not enough to counteract against cyber-threats. Hence, further research is needed in the area of human and computer collaboration against cyber-attacks.

## 5.2 Training and Awareness Raising

The cyber-security domain is built around a technological deterministic approach, where technology is assumed to be the remedy and only solution for every cyber-threat, while a more human-centered approach needs to be obtained. Concerning the high complexity of managing human factors, Nobles in [92] argued that information security experts should collaborate with cognitive scientists, human-factor specialists, and psychologists. According to Nobles, a collaboration like this could provide insight concerning the adoption of certain cyber-security behaviors by humans, thus presenting certain vulnerabilities. Apart from that, Nobles proposed several more approaches that would assist in decreasing the complexity of human-factor management:

- the sponsorship of research projects and training of employees on human factors
- the conduction of human-factor risk assessment
- the inclusion of human factors in the core of cyber-security practices and solutions
- the insertion of human factor courses in the certification process of cyber-security professionals
- the development of modern training and awareness-raising programs, such as gamification.

After interviewing seven cyber-security experts, J. Thomas in [114] managed to identify several factors that need to be taken into consideration in the assessment of cyber-security issues and user susceptibility:

- high-impact job roles,
- lack of information technology skills
- unfamiliarity with phishing attacks
- sophistication of attacks
- self-confidence level of the user
- user proficiency in phishing detection
- job roles that include email handling
- training and experience in cyber-security and information technology
- familiarity of the user with other phishing victims.

Practical approaches were also proposed in this research, such as dividing employees into distinct groups, according to the aforementioned factors, and developing training programs specifically for each group.

In a similar manner, Aldawood and Skinner in [3] suggested several innovative training approaches, such as:

- simulation methods to assess one's susceptibility,
- study of cyber-security issues independently through videos
- gamification to intrigue and draw attention in cyber-security,
- online methods (e.g., blogs about cyber-security) for synchronous and asynchronous interaction between users regarding cyber-security issues or incidents.

Additionally, the authors utilized a qualitative research method to study pitfalls and challenges of traditional and modern training and awareness-raising programs [4]. Both modern and traditional approaches face several issues, namely:

- *Business environmental* (e.g., prevailing technology, employee education)
- *Social* (e.g., influence of cultural characteristics)
- *Constitutional* (e.g., legal policies)
- *Organizational* (e.g., lack of personalized training and awareness programs)

- *Economic* (e.g., allocation of economic resources)
- *Personal* (e.g., personality traits, trusting human nature, lack of interest).

Furthermore, modern training techniques that include games and virtual labs, themed videos, simulations of real-life scenarios, etc. may be time-consuming and stressful for employees (due to high work pressure and urgency of deadlines). Another drawback of modern training techniques is that they may lack a simple design that can be understood by employees of non-IT background. They may also fail to address the vulnerabilities of each employee individually. On the other hand, traditional approaches are often described as boring and tedious, while they also lack exposure in real-life scenarios and critical evaluation of trust. Aldawood and Skinner also argued the necessity for the establishment of clear privacy and security policies by organizations, as well as for the conduction of penetration testing, prior to training and awareness-raising programs, to identify individual vulnerabilities and develop more personalized training techniques [4].

Table V HVA Frameworks

| Proposed by | Name | Approach | | | | |
|---|---|---|---|---|---|---|
| | | **HUMAN VULNERABILITY ASSESSMENT FRAMEWORKS** | | | | |
| [55] | Cyber Human Error Assessment Tool (CHEAT) | Assessment Domains | | | | |
| | | People | Organization | Technology | Environment | History |
| [1] | - | 17 cybersecurity scenarios | | | | |
| | | Modern Attacks | | Classical Attacks | | Attacks targeting children |
| [97] | Social Driven Vulnerability Assessment | 4 Phases | | | | |
| | | Information gathering about the participants | | Preparation of attack aproach (by utilizing the Victim Communication Stack - VCS) | Attack execution | Attack analysis |
| [2] | - | 4 principles that affect decision making | | | | |
| | | Socio-emotional | | Habitual | Perceptual | Socio psychological |
| [9] | - | 5 levels of evaluation | | | | |
| | | Definition | Data collection | Formulation | Representation | Attribution |
| [64] | - | Assessment Domains | | | | |
| | | Management | | Environment | Responsibility | Preparedness |

In [5] the same authors interviewed 21 cyber-security experts aiming to identify the most critical vulnerabilities in the cyber-security domain and investigate achievable mitigation strategies [5]. Most of the interviewees agreed that SE and human vulnerabilities pose the biggest threats in cyber-security. Moreover, they mentioned that lack of user awareness complicates things and renders cyber-defense more challenging. Countermeasures like traditional and non-traditional awareness training programs were proposed. Additionally, cyber-security professionals highlighted the necessity for continuous training, as well as for providing motivation to users for participating in such programs. Securing technical vulnerabilities and updating technical tools on a regular basis is assistive in the endeavor of enhancing cyber-security; however, it was clearly stated that organizations should invest in humans, while governments should enforce cyber-security laws, and schools should provide appropriate training in cyber-security issues.

In the context of raising phishing awareness, training programs would be more effective in cases where the simulated attack resembles real-life situations. Higashino in [62] proposed the design of a system that will allow organizations to share information about phishing emails. However, according to the author's proposal, no actual phishing emails will be shared due to privacy and security issues. On the contrary, only a part of the phishing email (after the removal of malicious URLs and personal information) will be uploaded to a phishing site server. The server will mimic the same environment of the actual attack and will be used for training purposes. The author highlighted the necessity of a simply structured training system for all organizations' employees to be able to sign in and attend the training program, even without explicit prior knowledge about information technology and security.

Several different awareness-raising programs and training approaches are being developed and adopted by the Information Technology industry for educating end-users about cyber-security issues and mitigation tactics. Some of these programs and approaches produce promising results, while the research community agrees that all training approaches should become more personalized to enhance effectiveness. Cuchta *et al.* in [38] explored human risk factors in cyber-security and investigated the results of three awareness-raising training methods. They conducted an experiment involving 4.777 staff, faculty members, and students from the Fairmont State University (West Virginia, United States). During a two-month period, 90.000 emails were sent to the participants, where, each day, emails contained a different deception approach to attract users to click on phishing URLs. The awareness-raising methods that were utilized were the following:

- *long, text-based documents*, where participants were presented with a 28-page document informing them about phishing
- *phishing email examples*, where the participants could observe certain highlighted keywords included in phishing emails, that should provoke suspicion
- *interactive games*, where participants were first trained on how to distinguish phishing URLs, and then they had to navigate through legitimate and malicious URLs. Finally, they had to decide about how to act. (i.e., click on the URL or not)

The results of this research revealed that almost half of the people recruited for this experiment (44.3%) fell victims to the simulated attacks, while 20% revealed their official university credentials. Regarding the awareness-raising method, the best outcome was observed when participants were presented with phishing email examples, which trained them to recognize suspicious vocabulary utilized in phishing emails. What is more, this approach was found to be even more effective when the examples were presented to the participants after they had already fallen victims to phishing.

From another point of view, Veksler *et al.* in [118] investigated the adoption of cognitive behavioral models in cyber-security, and the advantages that an approach like this may offer. Cognitive behavioral models describe how emotions, feelings, and behavior of individuals are triggered by spontaneous thoughts, which are shaped by one's beliefs and perceptions [17]. By using these models, human cognition and behavior can be simulated, regarding cyber-security defenders, network users, and cyber-attackers, wherein their interaction with actual environments can be monitored. The significance of consulting cognitive behavioral theory is highlighted in [118], regarding training programs design, real-time attack prediction, and cyber-security research. Regarding cyber-attack prediction, the authors explored the adoption of behavioral game theory, and argued its use towards better understanding attack patterns and attackers' behavior. In an earlier study [119], the authors had showcased that users' decisions can be predicted more accurately by a cognitive-based model., This prediction can also be assisted by the assumption that in real-life scenarios people neither follow a normative decision-making process – they usually rely on past experiences – nor a completely unpredictable approach. The authors also revealed that cognitive behavioral models can bypass the unpredictability of attackers' actions and decisions, because they can be adjusted dynamically, due to their ability of parameter fitting. Lastly in [118], the authors advocated that the use of cognitive models could provide significant aid to cyber-security defenders. Defenders usually experience heavy mental payload, due to the demanding security-related tasks that they perform. This attributes to human vulnerabilities caused by stress in the work environment, which were discussed earlier in this paper. Cognitive behavioral models could be utilized to provide computerized aid to the defenders when it is needed. This aid could assist defenders in performing their complicated tasks. It could also prevent defense performance from decreasing, due to heavy mental payload.

## 6 CONCLUSION

The aim of this paper was to investigate the existing literature related to human vulnerabilities and explore the reasons behind individual susceptibility. The research process led us to identify many different characteristics and situations that affect or shape human vulnerabilities, namely:

- personality traits
- cognitive processes, such as information-processing, decision-making, security-related behavior intentions, protection motivation, and risk-taking
- demographics like gender, age, and cultural background
- profession and computer experience
- current social situations, such as the outbreak of Covid-19
- workload, as well as stress and pressure in the work environment

Many people try to exploit these vulnerabilities to their advantage by utilizing several persuasion strategies. Lately, this is a current trend in the realm of cyber-security, where most malicious actors target humans instead of machines. The evolution and intensified use of social media, alongside the "naive" and trusting human nature, have brought a new kind of cyber-attackers to the fore, the SE attackers, who aim to initiate their attacks through social networks. Moreover, the professions that require the use of information technology, to some extent, have grown in number. Remote working has also increased dramatically since March 2020, due to the pandemic, thus monitoring the security conditions of the work/office environment has become more complex. All the aforementioned facts led to the expansion of the number of intrusion points that can be exploited by cyber-criminals.

Current HVA frameworks were studied during our research, as well as mitigation and prevention tactics towards user susceptibility. It was observed that state-of-the-art approaches in HVA lack a framework that gathers all the factors that influence human susceptibility – factors that were discussed in this paper and included in the proposed user susceptibility profile. Moreover, HVA should not be a one-time assessment but a continuous practice, since an individual may present different levels of susceptibility to cyber-threats each day (depending on time pressure, mental-load, mood, etc.). Additionally, HVA should not be a separate procedure, rather it should be included in the overall cyber-security capacity assessment of an organization. On top of that, current HVA frameworks lack the inclusion of user maliciousness assessment, which is considered to be a human-factor vulnerability. This may be explained by the fact that it is only recently that maliciousness – in the context of cyber-security – started to attract the researchers' attention. However, user maliciousness is recognized as an important variable in the cyber-security equation, as it can be utilized as a predictor of insider threats, and the inclusion of its assessment in an HVA framework is considered to be pivotal.

What is more, it is worth mentioning that many researchers in the field of information technology and cyber-security have focused their attention on phishing studies, in an endeavor to discover what affects susceptibility and what shapes human vulnerability. However, little research has been conducted regarding ethical implications of these studies. In addition to this, until recently, most of the studies that investigated the consequences of cyber-attacks dealt with the repercussions that firms face when falling victims to cyber-criminals. However, individual welfare is at stake too, and cyber-attacks' implications at a personal level are still neglected from recent studies.

Another theme that was explored in this paper is training and awareness-raising methodologies regarding cyber-security issues. Research upon human vulnerabilities has revealed that the necessity of preserving security demands a significant increase in end-user awareness towards cyber-security incidents. Users also need to be educated about cyber-security best practices. Furthermore, each human is susceptible to different persuasion methods and attack tactics, thus, traditional and modern training techniques need to evolve into more personalized approaches that will target and address the vulnerabilities of each individual. However, a variable that has not been adequately explored yet is the attackers' behavior. Realistic modelling of attacker and defender agents is an important factor in the development of modern training approaches in cyber-security. This is the case because simulations that resemble real-world scenarios would empower users to counteract cyber-attacks in real-life situations.

Conclusively, this paper presented several limitations of the current state-of-the-art regarding human vulnerabilities in cyber-security, through an extensive review of the literature. These identified limitations may present a starting point for new scientific research. It is our aspiration that this work will assist cyber-security researchers in their endeavor to identify future steps in the enhancement of cyber-security solutions.

# A. APPENDICES

## A.1 List of Used Acronyms

| Abbreviation | Definition |
|---|---|
| IoT | Internet of Things |
| IoE | Internet of Everything |
| SE | Social Engineering |
| HVA | Human Vulnerability Assessment |
| PVA | Personality Vulnerability Assessment |
| OCTAVE | Operationally Critical Threat Asset and Vulnerability Evaluation |
| FFM | Five Factor Model |
| BFI | Big Five Inventory |
| MBTI | Myers-Briggs Type Indicator |
| SD3 | Short Dark Triad |
| HSM | Heuristic Systematic Model |
| ELM | Elaboration Likelihood Model |
| SJT | Social Judgment Theory |
| GDMS | General Decision-Making Styles |
| DoSpeRT | Domain Specific Risk-Taking Scale |
| PRT | Passive Risk-Taking |
| PMT | Protection Motivation Theory |
| SeBIS | Security Behavior Intentions Scale |
| STPS | Susceptibility to Persuasion Scale |
| LSRP | Levenson Self Report Psychopathy Scale |
| PHIT | Phishing Internet Task |
| YDL | Yerkes-Dodson Law |
| CHEAT | Cyber Human Error Assessment Tool |
| SDVA | Social Driven Vulnerability Assessment |
| HAV | Human Attack Vector |
| VCS | Victim Communication Stack |
| HRA | Human Reliability Assessment |
| SCQ | Statistical Quality Control |
| HEART | Human Error Assessment and Reduction Technique |
| PDSA | Plan-Do-Study-Act Cycle |